%% file: MAIN-twocol-version.tex
\documentclass[amsmath,amssymb,superscriptaddress,longbibliography,twocolumn]{revtex4-2}
\usepackage[final]{pdfpages}
\usepackage{graphicx}
\usepackage{varioref}
\usepackage{xr-hyper}
\usepackage{nicefrac}
\usepackage{xfrac}
\usepackage{todonotes}
\usepackage{orcidlink}
\usepackage{hyperref}
\hypersetup{colorlinks,linkcolor=blue,urlcolor=teal,citecolor=orange}
\usepackage{ulem}
\usepackage{siunitx}
\usepackage{dcolumn}
\usepackage{bm}
\usepackage{xcolor}
\usepackage{alphabeta}
\usepackage{booktabs}

\newcommand{\lsco}{{La}$_{1.88}${Sr}$_{0.12}${CuO}$_4$}

\makeatletter
\AtBeginDocument{\let\LS@rot\@undefined}
\makeatother
\begin{document}





\title{Direct High-Magnetic-Field Coupling to Stripe Order in a Cuprate Superconductor }

\author{L. Martinelli~\orcidlink{0000-0003-4978-8006}}
\email{leonardo.martinelli@physik.uzh.ch}
\affiliation{Physik-Institut, Universit\"{a}t Z\"{u}rich, Winterthurerstrasse 
190, CH-8057 Z\"{u}rich, Switzerland}%

\author{I.~Biało~\orcidlink{0000-0003-3431-6102}}%
\affiliation{Physik-Institut, Universit\"{a}t Z\"{u}rich, Winterthurerstrasse 
190, CH-8057 Z\"{u}rich, Switzerland}%

\author{J.~Oppliger~\orcidlink{0000-0002-0712-4343}}%
\affiliation{Physik-Institut, Universit\"{a}t Z\"{u}rich, Winterthurerstrasse 
190, CH-8057 Z\"{u}rich, Switzerland}%

\author{J. Küspert~\orcidlink{0000-0002-2905-9992}}%
\affiliation{ESRF, The European Synchrotron, 71 Avenue des Martyrs, CS40220, 38043 Grenoble Cedex 9, France}

\author{O. Gerguri}%
\affiliation{Paul Scherrer Institut, Switzerland}

\author{S.~Brunner}
\affiliation{Physik-Institut, Universit\"{a}t Z\"{u}rich, Winterthurerstrasse 
190, CH-8057 Z\"{u}rich, Switzerland}%

\author{B. Eggert}
\affiliation{Faculty of Physics and Center for Nanointegration Duisburg–Essen (CENIDE), University of Duisburg–Essen, Duisburg, 47057, Germany}

\author{M.~H.~Fischer~\orcidlink{0000-0003-0810-6064}}
\affiliation{Physik-Institut, Universit\"{a}t Z\"{u}rich, Winterthurerstrasse 
190, CH-8057 Z\"{u}rich, Switzerland}%

\author{J. Geck~\orcidlink{0000-0002-2438-0672}}%
\affiliation{Institut für Festkörper- und Materialphysik, Technische Universität Dresden, 01069 Dresden, Germany}

\author{M. Rahn~\orcidlink{0000-0001-7403-8288}}%
\affiliation{EP VI, Center for Electronic Correlations and Magnetism, Institute of Physics, University of Augsburg, D-86159 Augsburg, Germany}

\author{E. Fogh~\orcidlink{0000-0001-8305-4466}}%
\affiliation{School of Natural Sciences, Technical University of Munich, D-85748 Garching, Germany}

\author{J. Choi~\orcidlink{0000-0003-4616-4345}}%
\affiliation{Physik-Institut, Universit\"{a}t Z\"{u}rich, Winterthurerstrasse 
190, CH-8057 Z\"{u}rich, Switzerland}
\affiliation{Department of Physics \& Astronomy, Seoul National University, Seoul 08826, Republic of Korea}

\author{A. Miyata}
\affiliation{Hochfeld-Magnetlabor Dresden (HLD-EMFL) and Würzburg-Dresden Cluster of Excellence ctd.qmat, Helmholtz-Zentrum Dresden-Rossendorf, 01328 Dresden, Germany}

\author{O. Prokhnenko~\orcidlink{0000-0002-5376-1765}}%
\affiliation{Helmholtz-Zentrum Berlin für Materialien und Energie, D-14109 Berlin, Germany}

\author{Z. Islam}%
\affiliation{Advanced Photon Source, Argonne National Laboratory, Argonne, Illinois 60439, USA}

\author{F.~Igoa Saldaña}
\affiliation{Deutsches Elektronen-Synchrotron DESY, Notkestra{\ss}e 85, 22607 Hamburg, Germany}

\author{M.~v.~Zimmermann~\orcidlink{0000-0002-9320-6846}}
\affiliation{Deutsches Elektronen-Synchrotron DESY, Notkestra{\ss}e 85, 22607 Hamburg, Germany}

\author{R. Nickel}
\affiliation{ESRF, The European Synchrotron, 71 Avenue des Martyrs, CS40220, 38043 Grenoble Cedex 9, France}

\author{K.~Kummer~\orcidlink{0000-0003-3044-7957}}
\affiliation{ESRF, The European Synchrotron, 71 Avenue des Martyrs, CS40220, 38043 Grenoble Cedex 9, France}

\author{N.~B.~Brookes~\orcidlink{0000-0002-1342-9530}}
\affiliation{ESRF, The European Synchrotron, 71 Avenue des Martyrs, CS40220, 38043 Grenoble Cedex 9, France}

\author{A.~Painganoor}
\affiliation{Department of Physics, Technical University of Denmark, DK-2800 Kongens Lyngby, Denmark}
\affiliation{Institut Laue-Langevin, 71 avenue des Martyrs, CS 20156, Grenoble, 38042 Cedex 9, France}

\author{P.~C.~Forino \orcidlink{0000-0002-0946-6486}}
\affiliation{Department of Physics, Technical University of Denmark, DK-2800 Kongens Lyngby, Denmark}

\author{R.~Toft-Petersen~\orcidlink{0000-0001-7638-3675}}
\affiliation{Department of Physics, Technical University of Denmark, DK-2800 Kongens Lyngby, Denmark}
\affiliation{European Spallation Source ERIC, P.O. Box 176, SE-221 00, Lund, Sweden}

\author{N.~B.~Christensen~\orcidlink{0000-0001-6443-2142}}
\affiliation{Department of Physics, Technical University of Denmark, DK-2800 Kongens Lyngby, Denmark}

\author{X.~Hong~\orcidlink{0000-0001-6219-2851}}
\affiliation{Physik-Institut, Universit\"{a}t Z\"{u}rich, Winterthurerstrasse 
190, CH-8057 Z\"{u}rich, Switzerland}%

\author{Q. Wang~\orcidlink{0000-0002-8741-7559}}
\affiliation{Department of Physics, The Chinese
University of Hong Kong, Shatin, Hong Kong, China}
\affiliation{State Key Laboratory of  Quantum Information Technologies and Materials, The Chinese University of Hong Kong, Shatin, Hong Kong, China}

\author{T.~Kurosawa \orcidlink{0000-0001-5065-6823}}
\affiliation{Department of Applied Physics, Hokkaido University, Sapporo 060-8628, Japan}

\author{N.~Momono}
\affiliation{Department of Physics, Hokkaido University - Sapporo 060-0810, 
Japan}
\affiliation{Department of Applied Sciences, Muroran Institute of Technology, Muroran 050-8585, Japan}

\author{M.~Oda}
\affiliation{Department of Physics, Hokkaido University - Sapporo 060-0810, 
Japan}

\author{D.~V.~Novikov}
\affiliation{Deutsches Elektronen-Synchrotron DESY, Notkestra{\ss}e 85, 22607 Hamburg, Germany}

\author{A.~Khadiev \orcidlink{0000-0001-7577-2855}}
\affiliation{Deutsches Elektronen-Synchrotron DESY, Notkestra{\ss}e 85, 22607 Hamburg, Germany}

\author{T. Herrmannsdoerfer}
\affiliation{Hochfeld-Magnetlabor Dresden (HLD-EMFL) and Würzburg-Dresden Cluster of Excellence ctd.qmat, Helmholtz-Zentrum Dresden-Rossendorf, 01328 Dresden, Germany}

\author{A. Kurtanidze \orcidlink{0009-0006-5543-780X}}
\affiliation{Hochfeld-Magnetlabor Dresden (HLD-EMFL) and Würzburg-Dresden Cluster of Excellence ctd.qmat, Helmholtz-Zentrum Dresden-Rossendorf, 01328 Dresden, Germany}
\affiliation{Institut für Festkörper- und Materialphysik, Technische Universität Dresden, 01069 Dresden, Germany}

\author{K. Ollefs}
\affiliation{Faculty of Physics and Center for Nanointegration Duisburg–Essen (CENIDE), University of Duisburg–Essen, Duisburg, 47057, Germany}

\author{Z. Konôpkov}
\affiliation{European XFEL GmbH, Holzkoppel 4, 22869 Schenefeld, Germany}

\author{M. Andrzejewski}
\affiliation{European XFEL GmbH, Holzkoppel 4, 22869 Schenefeld, Germany}

\author{M. Tang}
\affiliation{European XFEL GmbH, Holzkoppel 4, 22869 Schenefeld, Germany}

\author{U. Zastrau~\orcidlink{0000-0002-3575-4449}}
\affiliation{European XFEL GmbH, Holzkoppel 4, 22869 Schenefeld, Germany}

\author{A. Pelka \orcidlink{0009-0001-3308-5376}}
\affiliation{Hochfeld-Magnetlabor Dresden (HLD-EMFL) and Würzburg-Dresden Cluster of Excellence ctd.qmat, Helmholtz-Zentrum Dresden-Rossendorf, 01328 Dresden, Germany}

\author{H. Höppner \orcidlink{0009-0000-1929-5097}}
\affiliation{Hochfeld-Magnetlabor Dresden (HLD-EMFL) and Würzburg-Dresden Cluster of Excellence ctd.qmat, Helmholtz-Zentrum Dresden-Rossendorf, 01328 Dresden, Germany}

\author{J. S. Dambietz}
\affiliation{European XFEL GmbH, Holzkoppel 4, 22869 Schenefeld, Germany}
\author{V. Rovensky}
\affiliation{European XFEL GmbH, Holzkoppel 4, 22869 Schenefeld, Germany}

\author{T. Laurus \orcidlink{0000-0002-2258-2123}}
\affiliation{Deutsches Elektronen-Synchrotron DESY, Notkestra{\ss}e 85, 22607 Hamburg, Germany}

\author{E. Brambrink}
\affiliation{European XFEL GmbH, Holzkoppel 4, 22869 Schenefeld, Germany}

\author{B. Näser}
\affiliation{European XFEL GmbH, Holzkoppel 4, 22869 Schenefeld, Germany}

\author{M. Sikora \orcidlink{0000-0003-4491-3496}}
\affiliation{National Synchrotron Radiation Centre SOLARIS, Jagiellonian University, Czerwone Maki str. 98, 30-392 Kraków, Poland}

\author{C. Strohm~\orcidlink{0000-0001-6384-0259}}%
\email{cornelius.strohm@desy.de}
\affiliation{Deutsches Elektronen-Synchrotron DESY, Notkestra{\ss}e 85, 22607 Hamburg, Germany}

\author{C. Baehtz~\orcidlink{0000-0003-1480-511X}}%
\email{carsten.baehtz@xfel.eu}
\affiliation{European XFEL GmbH, Holzkoppel 4, 22869 Schenefeld, Germany}

\author{Sh. Yamamoto \orcidlink{0000-0001-5819-2867}}%
\email{s.yamamoto@hzdr.de}
\affiliation{Hochfeld-Magnetlabor Dresden (HLD-EMFL) and Würzburg-Dresden Cluster of Excellence ctd.qmat, Helmholtz-Zentrum Dresden-Rossendorf, 01328 Dresden, Germany}

\author{J.~Chang~\orcidlink{0000-0002-4655-1516}}
\email{johan.chang@physik.uzh.ch}
\affiliation{Physik-Institut, Universit\"{a}t Z\"{u}rich, Winterthurerstrasse 
190, CH-8057 Z\"{u}rich, Switzerland}%

\date{\today}


\maketitle


\textbf{Superconductivity in cuprates emerges out of a complex normal state that hosts density waves, pseudogap physics, and strange metal properties. Here, we access this normal state by synchronizing free-electron laser x-rays with high-magnetic-field pulses up to 44~T. We observe a linear increase in charge order amplitude and correlation length that persists far above the vortex melting transition. This behavior is incompatible with standard phase competition between charge order and superconductivity. By means of conventional hard x-ray diffraction and magnetostriction, we show that applied fields also enhance monoclinic lattice distortions. However, this magnetoelastic response is weaker and an epiphenomenon of the stripe order enhancement. Combined with recent observations of field-linear spin freezing, our results point to a direct coupling between magnetic field and the spin component of stripe order in the high-field normal state --- a mechanism independent of superconductivity suppression that has so far remained hidden from scattering probes.}\\

\input{TextInput_v3}

\medskip
\input{Methods}

\medskip
\input{acknowledgments}

\medskip
\bibliography{lqmr_leo, references}

\includepdf[pages={{},1,2,3,4,5,6,7,8,9,10,11}]{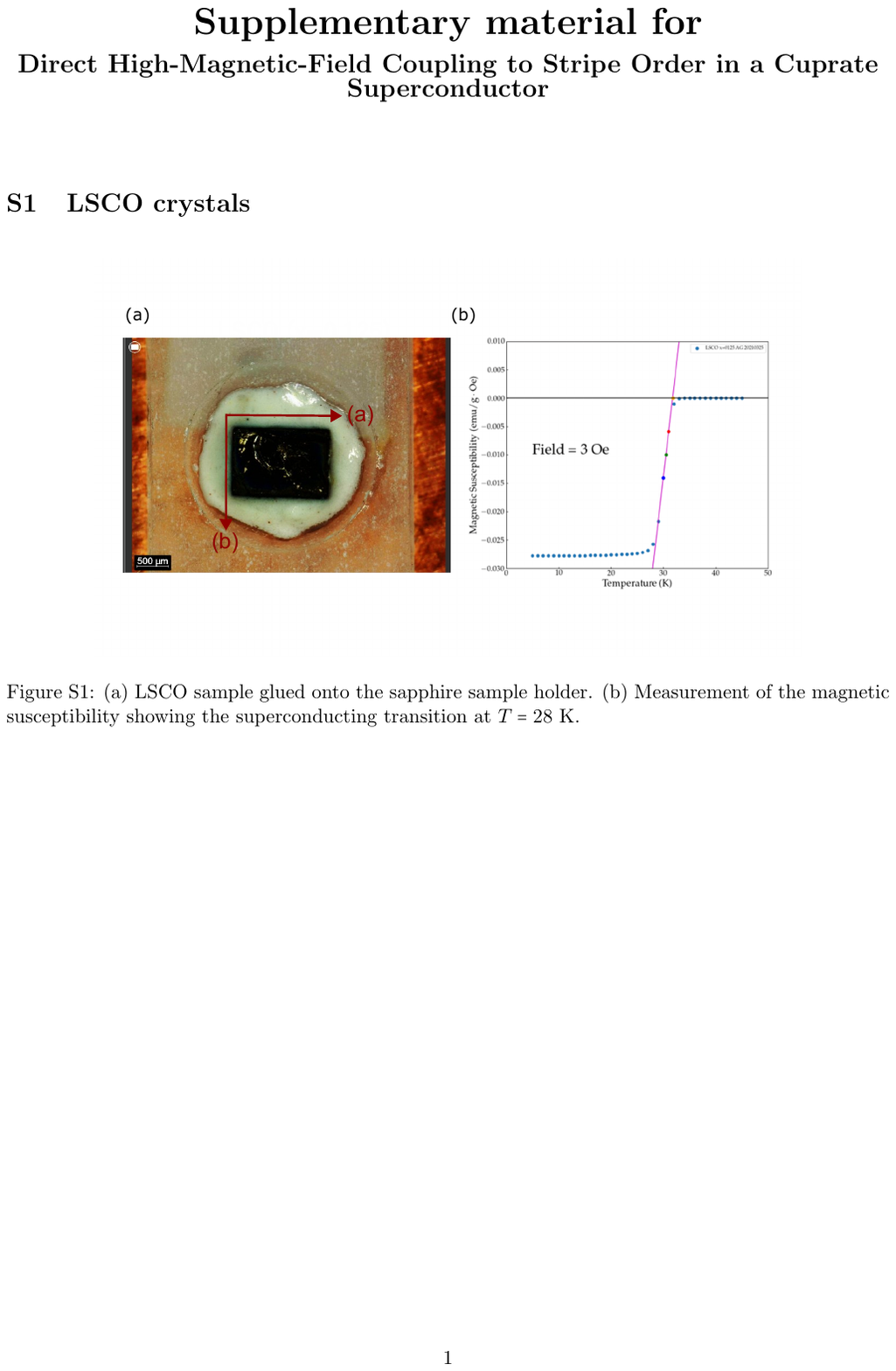}

\end{document}

%% file: TextInput_v3.tex

%
\begin{figure*}[t]
    \centering
    \includegraphics[width=0.999\textwidth]{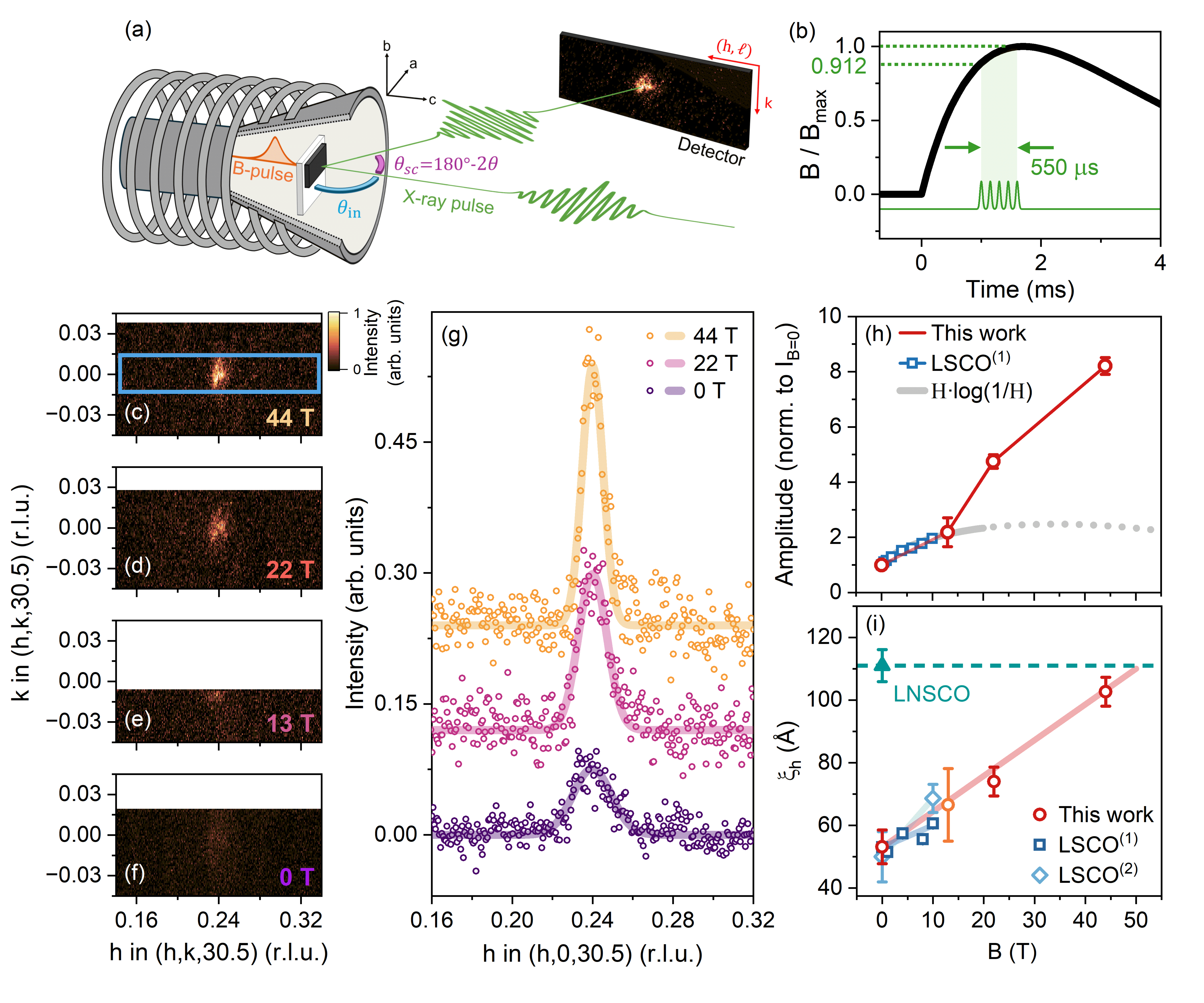}
    \caption{\textbf{Normal-state magnetic field effect on stripe order.} (a) Schematic illustration scattering geometry and the magnetic field sample environment. The sample sits inside a conical aperture with a longitudinal magnetic field directed perpendicular to the sample's surface along the crystallographic (001) direction. Magnetic field pulses are synchronized with the XFEL train. X-ray pulses impinge at an incident angle $\theta_\text{in}\!\sim\!66^{\circ}$ and are scattered on the detector at scattering angles $\theta_{\text{sc}}\!\sim\!44^{\circ}\pm2^{\circ}$, corresponding to the CO reflection $(0.24,0,30.5)$. (b) Sketch of the time trace of the magnetic field pulse (black solid line) and the XFEL x-ray train (green curve). The 
   first train pulse is synchronized with 91.2\% of maximum field. (c-f) Background subtracted intensity within the $(h,k,30.5)$ scattering plane for magnetic fields as indicated. A powder ring has been subtracted as explained in the Supplementary Material. Vertical movement of the peak stems from mechanical oscillations (see Methods section). (g) Diffracted intensity along the reciprocal $h$ direction through the charge ordering vector at $18$~K, obtained by integrating vertically over the blue rectangle shown in panel (c). Solid lines represent Gaussian fits. (h,i) Charge order amplitude (h) and correlation length (i) versus $c$-axis magnetic field. The fitting procedure was different for the 13 T image as explained in the Methods section. Experimental data measured in this work (red circles) are compared to Ref.~\cite{christensen_bulk_2014} (LSCO$^\text{(1)}$, blue squares) and Choi et al.~\cite{choi_unveiling_2022} (LSCO$^\text{(2)}$, light blue diamonds). La$_{1.48}$Nd$_{0.4}$Sr$_{0.12}$CuO$_4$ (LNSCO) data are taken from Ref.~\cite{wilkins_comparison_2011}. Amplitude is plotted against  $\mathcal{H}\log (1/\mathcal{H})$ with $\mathcal{H}=B/B_{c2}$~\cite{demler_spinordering_2001}. Correlation length has been calculated for all datasets as the inverse of the half-width-at-half-maximum of the charge order reflection along the $h$ reciprocal space axis.}
    \label{fig:xfel_fig}
\end{figure*}
\begin{figure*}
    \center{\includegraphics[width=0.999\textwidth]{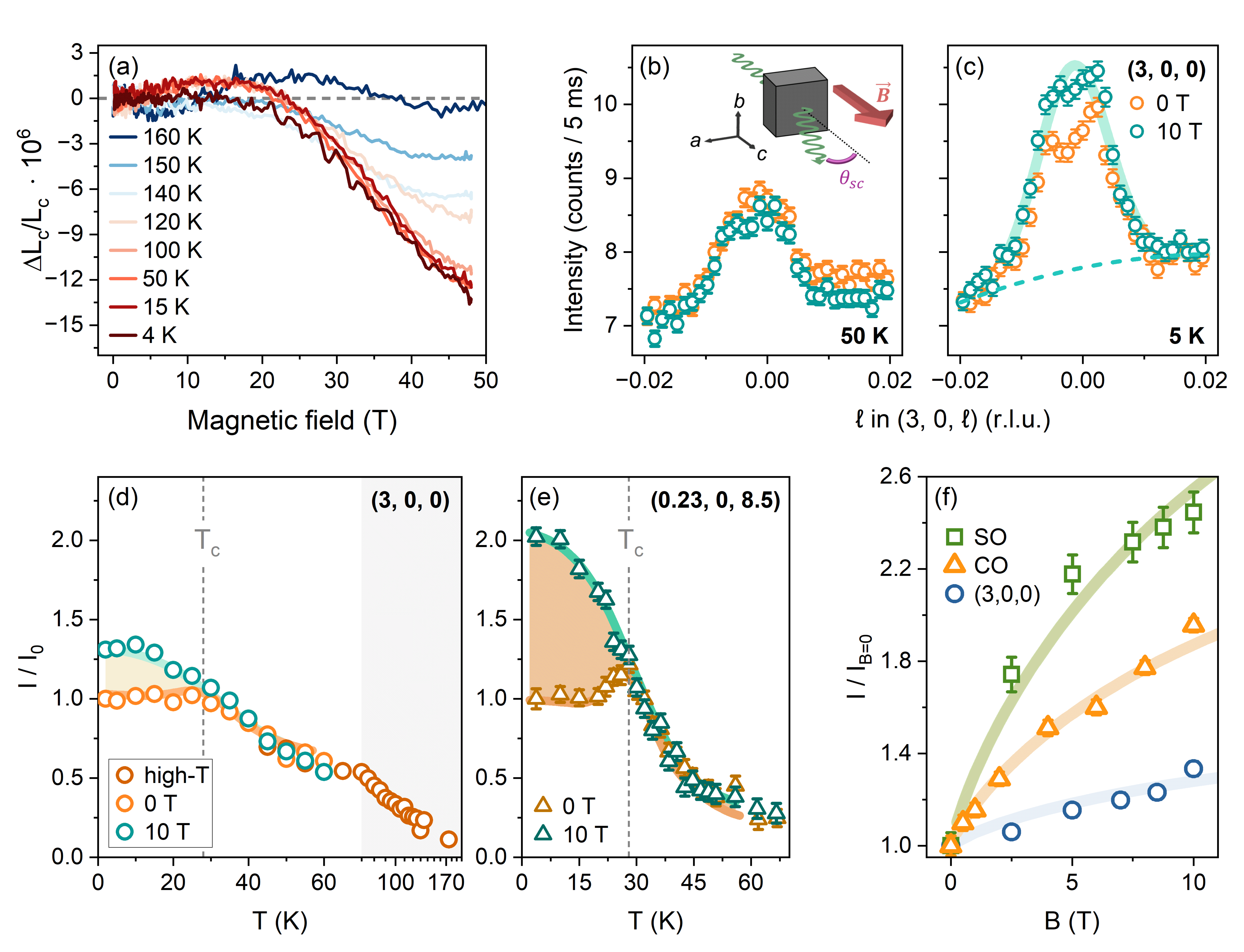}} 
    \caption{\textbf{Phase competition between monoclinic order and superconductivity.} (a) Out-of-plane magnetostriction in stripe ordered {La}$_{1.88}${Sr}$_{0.12}${CuO}$_4$. The $c$-axis lattice parameter change $\Delta c$ is plotted as $\Delta L_c/L_c=\Delta c/c$ as a function of temperature and $c$-axis magnetic field. Upon cooling a gradually larger magnetostriction effect is observed in high magnetic fields (b)-(c) One-dimensional  q-scans ($\ell$-scan) through the $(3 0 0)$ forbidden Bragg reflection as a function of magnetic field, for temperatures as indicated. Inset shows a sketch of the scattering geometry and sample environment. Incident beam and magnetic field are aligned along the crystallographic $c$-axis direction. This geometry yields a scattering vector along the pseudo-tetragonal $a$-axis direction.  (d-e) Temperature dependence of the  forbidden Bragg peak $(3,0,0)$ (d) and charge stripe order reflection $(0.24,0,8.5)$ (e) normalized to the lowest-temperature zero-field value. Data in panel (e) have been taken from Ref.~\cite{christensen_bulk_2014}. Both reflections grow upon cooling. In zero magnetic field, they  saturate when entering into the superconducting state. They display a clear magnetic field enhancement below $T_c$. In panel (d), the shaded region above $70\,$K is plotted in semilog scale (hence the apparent change in slope). (f) Magnetic field dependence of spin order (SO), green) at 2~K (from Ref.~\cite{chang_tuning_2008}), charge order (yellow triangles) at $3\,$K, and of the forbidden $(3,0,0)$ at $5\,$K (blue circles). Solid lines indicate the predicted dependence emerging from phase competition~\cite{demler_spinordering_2001}.}
    \label{fig:forbidden_bragg}
\end{figure*}

\noindent The ground state from which high-temperature superconductivity emerges in cuprates remains enigmatic~\cite{phillips2022Strangera}. Strong electronic correlations give rise to non-Fermi liquid behaviour~\cite{grissonnanche_linearin_2021} and phases with broken symmetries~\cite{ghiringhelli_longrange_2012}.
As the entire cuprate phase diagram is interaction driven, there is a strong interest in determining how the symmetry breaking couples to external fields.
Generally, symmetry breaking phases compete with superconductivity~\cite{fradkin_colloquium_2015}, which manifests itself through magnetic field effects inside the superconducting state~\cite{chang_direct_2012, ghiringhelli_longrange_2012}.
Magnetic field weakens
superconductivity, which in turn enhances the competing orders.
More recent measurements, however, suggest that the relationship between the two orders is more complex.
Recent high-field diffraction studies linked charge order (CO) field-induced effects to the melting of the vortex lattice \cite{wen_enhanced_2023}. This observation suggests that charge order is competing with a superconducting state characterized by long-range phase coherence, but might be compatible with local superconducting pairing~\cite{wen_observation_2019, lee2026Superconductivity}.
Finally, ultrasound and magnetotransport measurements of spin freezing showed an anomalous high-field response above the vortex melting transition~\cite{vinograd_competition_2022, frachet2021High}, difficult to reconcile with a direct competition with superconductivity. As such, there is no unified view of the magnetic field effects in the high-field phase of cuprate superconductors. 

Scattering and diffraction experiments provide direct access to the properties of broken symmetries. 
However, a straightforward application of these techniques with static magnetic fields is limited to 20~T. 
The advent of x-ray free electron lasers (XFELs) has opened new opportunities~\cite{doiron-leyraud_quantum_2007,yu_magnetic_2016, bortel_3d_2024}.
Extremely high brilliance and femtosecond temporal resolution enable the study of matter at extreme conditions such as planetary pressures or high magnetic fields. Indeed, single-shot experiments, where intense magnetic field pulses are synchronized with femto-second x-ray pulses, emerge as a new technique to probe enigmatic phases of matter~\cite{gerber_threedimensional_2015,jang_ideal_2016,ikeda_xray_2025,ikeda_single_2020,gen_xrd_2025}.
To address this topic, we combine x-ray free-electron laser measurements with a new high-field cryo-magnet. We investigate the fate of charge order in the high magnetic field phase of La$_{1.875}$Sr$_{0.125}$CuO$_4$ (LSCO). By studying the region of the phase diagram across the vortex melting transition $B_m$ and the upper critical field $B_{c2}$, we find no saturation of charge order enhancement and correlation length above $B_m$. 
To rationalize our results, we perform additional hard x-ray diffraction and magnetostriction measurements up to 50\,T. \\[1.5mm]
\subsection*{Giant field enhancement of stripe order:}
We performed pulsed-field x-ray diffraction exploiting free-electron laser radiation.
A sketch of our setup is illustrated in Fig.~\ref{fig:xfel_fig}a (see also the Methods section and the Supplementary Material).
Our cryo-magnet enables pulsed magnetic fields close to the incident x-ray beam direction, and allows the use of back-scattering geometry.
We apply magnetic fields along the crystallographic $c$-axis and access to intense charge order reflections. More importantly, synchronization of x-ray and intense magnetic (44~T) field pulses in combination with a high-speed two-dimensional detector enables single-shot diffraction. The detector frame probes essentially the $(h,k)$ reciprocal space plane.
Results are reported in Fig.~\ref{fig:xfel_fig}. Intensity in the $(h,k,30.5)$ scattering plane is shown in Fig.~\ref{fig:xfel_fig}(c-f) for magnetic fields as indicated. 
A significant magnetic-field enhancement of charge order is evident.
By integrating the intensity into a one-dimensional $q$-scan (Fig.~\ref{fig:xfel_fig}g), we extracted the charge stripe order peak amplitudes (Fig.~\ref{fig:xfel_fig}(h)) and correlation lengths (Fig.~\ref{fig:xfel_fig}(i)). Interestingly, no saturation of the field effect is observed.  For zero-field, the observed correlation length is consistent with existing literature~\cite{christensen_bulk_2014,wang_uniaxial_2022,choi_unveiling_2022}. 

As reported by resonant diffraction, the charge order structure rotates from the principal crystallographic axis with a small buckling angle~\cite{wang_uniaxial_2022}. This leads to broadening of the charge order diffraction peak in the direction perpendicular to the Cu-O bond direction~\cite{thampy_rotated_2014,wang_uniaxial_2022}. These results are reproduced in our XFEL experiment -- see Fig.~\ref{fig:xfel_fig}g. 
We characterized our LSCO $x=0.125$ crystals with resonant and non-resonant synchrotron scattering. Previously reported magnetic field effects~\cite{chang_magneticfieldinduced_2009,hucker_enhanced_2013,wen_enhanced_2023,lake_antiferromagnetic_2002} inside the superconducting state were reproduced using resonant inelastic x-ray scattering (RIXS) at the Copper $L_3$ edge up to 9 T (see Sec.~S6 in the Supplementary Material). In this low-field limit ($B<15$~T), the charge order reflection intensity scales approximately as expected from a phase competition scenario~\cite{demler_spinordering_2001}, as shown in Fig.~\ref{fig:xfel_fig}(h) and \ref{fig:forbidden_bragg}(f). \\[1.5mm]

Our key observation is that the charge component of stripe order field effect persists up to 44~T. Both the peak amplitude and the correlation length scale approximately linearly with magnetic field. This marks a clear deviation from the logarithmic saturation expected within phenomenological phase competition models~\cite{demler_spinordering_2001}. The amplitude is enhanced by a factor of eight and the correlation length doubles by application of 44~T. Interestingly, the correlation length approaches that of La$_{1.48-x}$Nd$_{0.4}$Sr$_x$CuO$_4$, a standard stripe-ordered compound where superconductivity is essentially absent around $x=0.125$~\cite{wilkins_comparison_2011}.\\[1.5mm]

\begin{figure}[t]
    \centering
    \includegraphics[width=0.49\textwidth]{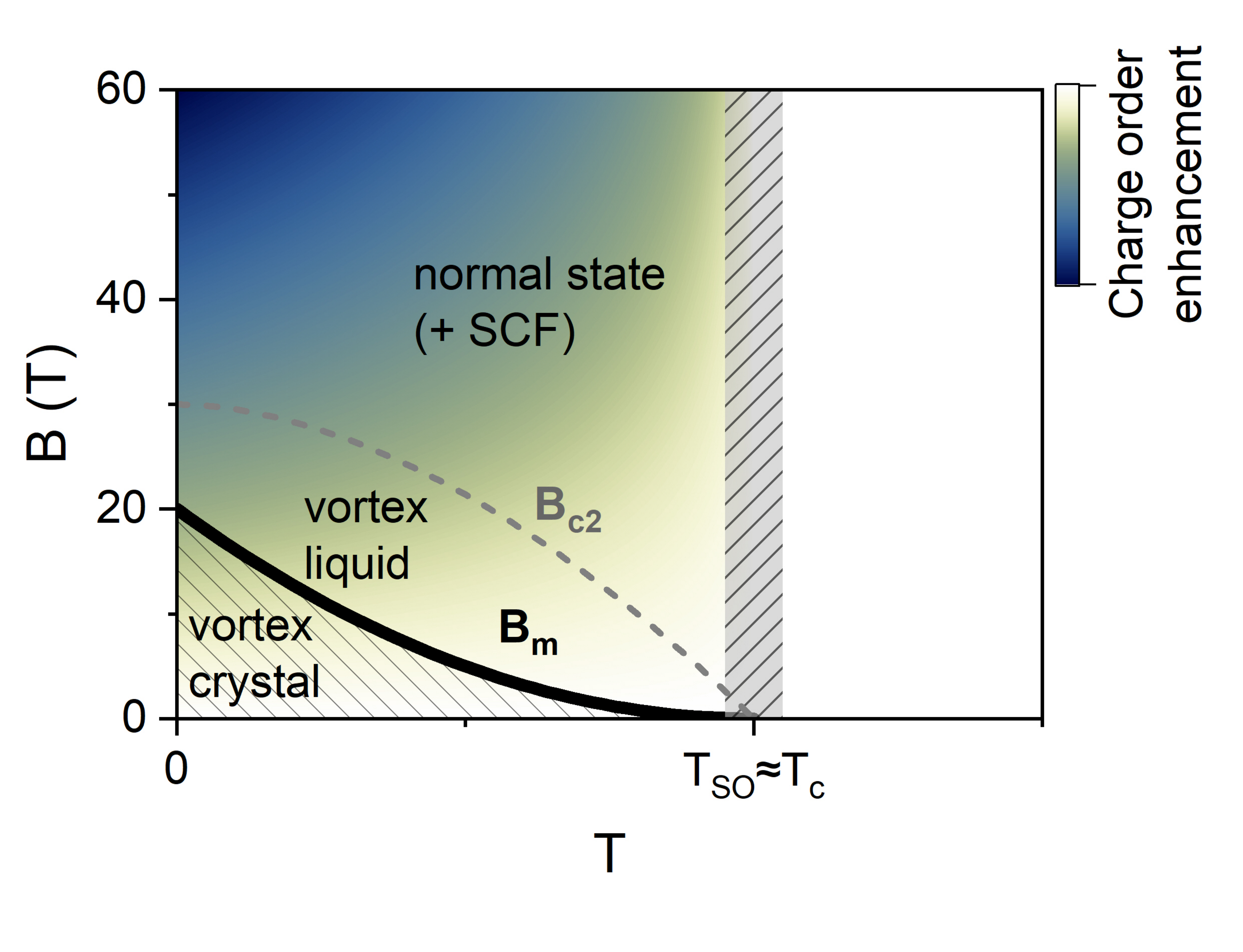}
    \caption{\textbf{Sketch of the charge order enhancement in the (T,B) phase diagram of {La}$_{1.88}${Sr}$_{0.12}${CuO}$_4$.} The enhancement is reported with a shaded color map. It increases linearly with field, and is present only below $T_c$~\cite{wen_enhanced_2023, hucker_enhanced_2013} $B_m$ and $B_{c2}$ denote the vortex melting transition and upper critical field, respectively. While the value of $B_m$ is precisely extracted from magnetotransport measurements~\cite{frachet2021High}, the value of $B_{c2}$ is reported with a dashed line to highlight that its value is not precisely measurable and might be just a crossover. The $T_\text{SO}\approx T_c$ line is drawn with a patterned rectangle to indicate that the two temperatures might not exactly coincide.}
    \label{fig:phase_diagram}
\end{figure}

\subsection*{Superconducting magnetic field scales:}
Superconductivity is associated with two magnetic field scales: 
the vortex melting field $B_m$ and the upper critical field $B_{c2}$.
The vortex melting transition breaks superconducting long-range phase coherence. In LSCO, magnetotransport measurements place it below $10\,$T~\cite{frachet2021High} at $T\sim15\,$K, well below the maximum field used in this experiment. A recent pulsed-field diffraction experiment has reported an enhancement of the charge order intensity above $B_m$~\cite{wen_enhanced_2023}, which is confirmed in our measurements as a clear deviation from the (low-field) $\log$-behaviour (see Fig.~\ref{fig:xfel_fig}(h)). In Ref.~\cite{wen_enhanced_2023}, the intensity increase is interpreted as a possible coexistence of short-range charge density waves and local superconducting pairing.
Our diffraction experiments however reveal 
the absence of saturation above the melting transition, excluding
the interpretation in terms 
{coexistence with short-range superconducting correlations.}


In the high-field cuprate phase diagram there is controversy on the upper critical field $B_{c2}$~\cite{wang_dependence_2003,chang_decrease_2012, frachet2020Hidden}.
This ill-defined field scale marks in essence the transition from vortex liquid into superconducting fluctuations in a normal state~\cite{blatter1994Vortices}. Transport and specific heat experiments on underdoped LSCO lead to very different estimations of $B_{c2}$. At around 1/8 doping, estimates range between 20-70~T at 0~K~\cite{chang_decrease_2012, wang2008Doping, frachet2020Hidden}, with more evidence towards the lower limit. At $18$~K, $B_{c2}$ should roughly be half of these values. 
In the accepted phase competition scenario, the magnetic field weakens superconductivity in the region between vortices, whose density increases linearly with field~\cite{demler_spinordering_2001}. This yields a $(B/B_{c2})\log(B_{c2}/B)$ dependence that saturates at high field. 
Reproducing a quasi-linear field effect up to 44~T within this picture would require an upper critical field of hundreds of Tesla (see Sec.~7 in the Supplementary Material).
We deem this unrealistically high and therefore assume that, above $25$~T,  of our experiment is carried out in a normal state hosting superconducting fluctuations.




A phase competition above the upper critical field is not entirely unexpected due to the presence of short-lived fluctuating Cooper pairs up to $B\gg B_{c2}$. The existence of such fluctuations has been proven in Nernst effect measurements of several type-II superconductors~\cite{pourret2006Observation, pourret2007Length}, including cuprates~\cite{chang_decrease_2012}. However, in other cuprates superconducting fluctuations compete less strongly with charge order than the long-range condensate~\cite{wu_emergence_2013}, and provide no quantitative mechanism for the observed enhancement. 

Phase competition with long-range superconductivity gives unreasonable values for the critical field, and effects due to vortex physics and normal-state fluctuations cannot account for the observed field dependence. These considerations point to a direct coupling between the magnetic field and either lattice or spin degrees of freedom.\\[1.5mm]

\subsection*{Magnetoelastics and strain-waves:}
%
We investigate possible direct interactions with the lattice through magnetoelastic measurements.
A direct coupling between stripe order and the lattice could explain a normal-state enhancement of charge order in the presence of a magnetoelastic effect.
We therefore probed $c$-axis magnetostriction $\Delta L_c/L_c$ of LSCO $x=0.125$ in a wide range of temperatures and magnetic fields. At high temperature ($T>150$~K) the magnetostriction effect is vanishingly small (see Fig.~\ref{fig:forbidden_bragg}(a)). A gradually stronger magnetostriction effect is found upon cooling deeper into the orthorhombic phase. At base temperature ($4$~K)  $\Delta L_c/L_c$ scales approximately with $B^2$
and reaches $\sim\!1.3\cdot10^{-5}$ at 48~T. 
The value of the induced $c$-axis strain is relatively modest, two orders of magnitude smaller than what can be achieved in uniaxial strain experiments~\cite{kuspert_engineering_2024, bialo2026Orbital, martinelli_decoupling_2025}. While such strain may weakly affect orthorhombic and monoclinic distortions, it is far too small to account for the dramatic enhancement of stripe order.

The charge component of stripe order in LSCO is symmetry-incompatible with either tetragonal or orthorhombic crystal structures. Monoclinic distortions are therefore either a prerequisite or an integral part of stripe order~\cite{frison_crystal_2022}. In the pseudo-tetragonal notation, monoclinic strain waves manifest by (forbidden) Bragg reflections at $(h,0,0)$ and $(0,0,\ell)$, with $h,\ell$ being odd integers. 

We conducted a comprehensive study of the forbidden Bragg peaks as a function of temperature and magnetic field (see Fig.~\ref{fig:forbidden_bragg}(b-d)). In the normal state ($T>T_c\approx30$~K), monoclinic distortions grow upon cooling (see Fig.~\ref{fig:forbidden_bragg}d).
Around the superconducting transition temperature, however, the monoclinic order saturates in zero magnetic field (Fig.~\ref{fig:forbidden_bragg}f), exhibiting the same qualitative behaviour as charge order.
For moderate magnetic fields ($<10\,$T) the monoclinic order is enhanced, but only 
below $\approx30$~K.
The effect 
increases with magnetic field (Fig.~\ref{fig:forbidden_bragg}f) -- similar (but three and five times smaller, respectively) to what has been reported for charge and spin components of stripe order~\cite{christensen_bulk_2014, chang_tuning_2008, kuspert_engineering_2024}.
These observations establish a direct coupling between monoclinic strain waves and spin/charge stripe order.

\bigskip


\subsection*{Discussion:}
High-brilliance light pulses might by themselves interact and alter quantum ground states. In the cuprates and related materials, there are reports on light-induced transient superconductivity~\cite{fava_magnetic_2024,fausti_lightinduced_2011,budden_evidence_2021}. Light-quenching effects have also been reported in the context of charge order competition with superconductivity in the cuprate materials~\cite{jang_characterization_2022,wandel_enhanced_2022}.
We however attribute the reported observations primarily to the external magnetic field, as light pulses tend to weaken rather than enhance the charge order parameter~\cite{mitrano_ultrafast_2019}.

The observation of linearly magnetic-field enhanced charge order in LSCO is difficult to reconcile with phase competition. A linear increase in the charge order intensity is expected in the superconductive phase, as the density of vortices increases linearly with field (see Sec.~7 in the Supplementary Material). However, as argued above, it would either require an unreasonably large upper critical field or unexpected interactions between short-range superconducting fluctuations and charge order. 
Having ruled out phase competition, we now assess the relative roles of lattice, charge, and spin degrees of freedom.

The weak magnetoelastic effects suggest that the lattice plays a subordinate role.
First, the low-field effect on monoclinic distortions is three  times stronger on the charge component of stripe order, and even stronger on the spin sector.
Second, the high-field magnetostriction produces marginal strain effects, three orders of magnitude smaller than what has been achieved in uniaxial strain experiments~\cite{ kuspert_engineering_2024, bialo2026Orbital}. Third, the high-field stripe order enhancement is only present below the onset of spin order and superconductivity, consistent with moderate fields results~\cite{choi_unveiling_2022, christensen_bulk_2014, kuspert_engineering_2024, wen_enhanced_2023} (see also Sec.~S5).
A leading lattice effect should not display such a discontinuity at $T_\text{SO}\approx T_c$, given that the monoclinic distortions persist to much higher temperatures.

At the same time, a direct, giant high-field enhancement of the charge component of stripe order is likely a secondary consequence of a coupling to spin. In the low-field limit the field effect on the spin is stronger than that on the charge, as shown in Fig.~\ref{fig:forbidden_bragg}. Moreover, the 30~K onset of the field enhancement indeed coincides not only with $T_c$, but also with the spin order onset temperature $T_\text{SO}$. We therefore conclude in favor of a magnetic field coupling to spin stripe order.
Our interpretation aligns with  recent NMR and ultrasound experiments in high magnetic fields, which reported an unusual response of the spin sector~\cite{frachet2021High, vinograd_competition_2022}. The spin-freezing energy scale, related to the spin stiffness, increases linearly with field, with no saturation up to 60~T. Similarly, the magneto-elastic coupling also monotonically increases. Interestingly, the ordered magnetic moment instead only shows a mild dependence with field~\cite{chang_tuning_2008}. 

Our results provide a direct evidence that the response of charge (and by comparison with recent work, spin)  order to magnetic field is not merely a consequence of superconductivity suppression. 
Above the vortex melting transition, the field couples directly to the spin component of stripe order with a mechanism that persists in the low-temperature, high-field normal state.
Access to this region has until now been limited to transport, NMR and thermodynamic probes, outside the reach of scattering experiments. The present work closes this gap and demonstrates that current FEL and pulsed-field technology can enable a new class of experiments on cuprates and other correlated-electron systems.
Our findings add a direct, normal-state coupling channel to the set of interactions shaping the high-field cuprate phase diagram — one that must be disentangled from phase competition in any comprehensive description of intertwined orders.



%% file: Methods.tex
\section*{Methods}\label{sec:methods}


\textit{Characterization of samples and charge order reflection:} A single crystal rod of La$_{2-x}$Sr$_x$CuO$_4$ with $x=0.125$ was grown by floating zone methodology~\cite{chang_tuning_2008}. A rectangular piece was cut with dimension 2mm x 1mm x 0.5mm along the $a$,$b$ and $c$ axes, respectively, and aligned using Laue diffraction. The (001) was normal to the sample surface. The sample was then glued onto a monocrystalline sapphire sample holder using a shallow layer of silver epoxy for thermal contact on the bottom, and Torr Seal on the sides (see Fig.S1).
The superconducting transition inferred from magnetic susceptibility is $T_c=30\pm1$~K, with a sharp transition indicative of low disorder.
Prior to the pulsed magnetic field experiments, our LSCO samples were characterized for charge order by both resonant and non-resonant x-ray scattering. The non-resonant characterization was carried out at the P23 beamline at the PETRA-III synchrotron - DESY Hamburg. We used the same experimental geometry and energy ($15.5$~keV) of the XFEL experiment, with the difference that the scattering plane was vertical. A LAMBDA detector with GaAs sensor was employed. We identified several charge order reflections at $Q=(\delta,0,\ell+0.5)$ with $\delta=0.23$ and $\ell=26,28,30$ reciprocal lattice units (r.~l.~u.) and characterized them as a function of temperature. A more detailed characterization of the charge order reflections, including the temperature dependence, is given in Sec.~S3.

\medskip
\textit{Resonant soft x-ray scattering in static magnetic fields:} 
To characterize the magnetic-field enhancement of charge order in moderate fields, we performed Resonant Inelastic X-ray Scattering measurements. The experiments were carried out on the ID32 beamline of the European Synchrotron Radiation Facility (ESRF)~\cite{brookes_beamline_2018}. The cryomagnet allowed magnetic fields up to 9~T. The measurements were performed at the copper $L$-edge ($931\,$eV), and the energy resolution ($310\,$meV) was determined by measuring amorphous silver paint on the sample holder. Scattering angle was fixed to $\theta_\text{sc}=90$° and momentum scans were acquired by rocking the incident angle $\theta$. We studied the $(0.24,0)$ charge order reflection at $L\sim1.2\,$~r.l.u.. The magnetic field was applied along the x-ray beam. In the grazing-outgoing configuration used for the measurements, it was at an angle of $8$° from the sample $c$-axis. Temperature was fixed to $4$~K. The results and more details are presented in the Supplemental Material.


\medskip
\textit{Non-resonant, high-energy x-ray diffraction in static magnetic fields:} Non-resonant diffraction experiments in static magnetic fields up to 10 Tesla were carried out at the P21.1 beam line at the PETRA-III synchrotron. We used a 10~T cryomagnet in the temperature range $2-300$~K. We employed $101.4$~keV photons and a transmission geometry. Our single crystal was aligned with the pseudo-tetragonal crystallographic $a$- and $c$-axes in the horizontal scattering plane with magnetic field applied along the $c$-axis. Scattered photons were recorded using a Dectris Pilatus 100K CdTe detector.

\medskip
\textit{Non-resonant, x-ray diffraction FEL experiment:} Pulsed-field experiments at the European Free electron laser facility were carried out at the HED beamline. 
The electron beam energy was 16.3~GeV.
We used 15.5~keV SASE radiation with a bandwidth
$\Delta E/E \approx 10^{-3}$  in a back-scattering geometry. The magnetic field pulse was generated by a liquid-nitrogen-cooled magnet with a $20-60$° biconical opening and measurements were conducted across the $60$° opening. The magnet was immersed in liquid nitrogen during operation. The capacitor bank was operated at a charging voltage of 16~kV for the maximum field pulse, 50.4~T, corresponding to a stored energy of approximately 332.8~kJ. At the sample position close to the exit of the conical opening, the magnetic field has a value $(94\pm1)\%$ of the maximum field. To synchronize the x-ray probe with the pulsed magnetic field, the European XFEL was operated in the long-pulse-train mode, in which every 100\textsuperscript{th} train was delivered exclusively to the HED station. In this mode, the pulse-train duration was 550~$\text{\mu}$s. The timing was adjusted such that the last x-ray pulse in the train coincided with the maximum of the magnetic-field pulse. Accordingly, the first x-ray pulse in the train probed the sample at a field corresponding to 91.2\% of the maximum field (see Supplementary Material).  
We accessed a portion of reciprocal space $(h,0,\ell)$ with $\ell\gg h$. Specifically, we probed the $(0,0,30)$ and $(2,0,32)$ Bragg reflections for alignment and the $(\delta,0,30.5)$ charge stripe order reflection (see Fig.~S2 in the Supplementary Material).
We employed a Stinger cold-finger cryostat, and the measurements were conducted at the base temperature of 18~K. We used an Adaptive Gain Integrating Pixel (AGIPD) SPARTA detector, capable of imaging single x-ray pulses and synchronized with the XFEL time structure. More information regarding the experimental setup is given in Sec.~S2 of the Supplementary Material.
The vertical movement of the peak on the detetor is caused by a mechanical vibration of the sample stage during application of magnetic fields. The intensity of the peak remains constant and no change in scattering angle is observed.
The 13~T point in Fig.~\ref{fig:xfel_fig}(i) (reported in orange) sits partially outside the detectore frame. The fitting was performed by fixing the vertical width to the value obtained by interpolating between the 0~T and 22~T measurements. 

\medskip
\textit{Magnetostriction experiments in pulsed field:}\label{subsec:magneto}
The longitudinal magnetostriction measurements were performed using the fiber Bragg grating method~\cite{daou2010High} in pulsed magnetic fields parallel to the \textit{c}-axis at the High Magnetic Field Laboratory (HLD) at the Helmholtz-Zentrum Dresden-Rossendorf (HZDR). The relative length change along the $c$-axis, $\Delta c/c$, was determined from the shift in the Bragg wavelength.


\begin{table}[ht]
\caption{\textbf{Experiments.} Overview of instruments, photon energy, momentum, temperature and magnetic field conditions under which monoclinic and charge stripe order is probed in \lsco. Magnetic field applied along the crystallographic $c$-axis.}
\label{tab:tab1}
\begin{tabular}{@{}ccccc@{}}
\toprule
Beam line & $E$ (keV) & $Q$ (r.l.u.) & $T$ (K) & Field (T) \\
\midrule
P23   & 15.5 & ($\delta$,0,28.5$\rightarrow$30.5) & 10--100 & 0 \\
ID32  & 0.93 & ($\delta$,0,1.5)                   & 4  & 0--9 \\
HED   & 15.5 & ($\delta$,0,30.5)                  & 18--60  & 0--50 \\
\midrule
P21.1 & $101.4$  & (3,0,0)                             & 2--200  & 0--10 \\
\botrule
\end{tabular}
\end{table}

%% file: acknowledgments.tex
\textit{Author contributions:}
Single crystals were grown by T.K., N.M., and M.O.. The XFEL experiment were carried out by L.M., I.B., J.O.,J.K., J.G., M.R., E.F., O.G., O.P., Z.I., N.B.C., B.E., M.S., T.H., Z.K., M.A., M.T., U.Z., A.P., H.H., C.S., R.T.P., C.B.,S.Y., and J.C.. During this experiment, the Sparta detector operation was supported by J.S.D, V.R., T.L. and timing system and pulser setup were operated by E.B. and B.N.. Resonant experiments at ESRF were carried out by L.M., I.B., K.K., and N.B.B.. Hard x-ray experiments at PETRA-III (P021), were carried out by L.M., I.B., S.B., J.O., X.H., Q.W., F.I.S., M.v.Z., and J.C. Magnetostriction experiments were carried out by Jaewon C. and A.M.. Preparatory x-ray experiments at PETRA-III (P023) were carried out by L.M, I.B., J.O., O.G., D.V.M., and A.K.. All sample preparation was carried out by L.M. and I.B.. Data analysis were carried out by L.M. (XFEL data), S.B., J.O. X.H. (PETRA-III) and Jaewon C. (magnetostriction). The manuscript was written by L.M. and J.C. with input from all authors. 

\textit{Acknowledgments:} Insightful discussions with David Leboeuf and Marc-Henri Julien are acknowledged. We also acknowledge the European XFEL in Schenefeld, Germany, for provision of X-ray free electron laser beam time at the Scientific Instrument HED (High Energy Density Science) under proposal number 6776 and thank the staff for their assistance. The authors are indebted to the HIBEF user consortium for the provision of instrumentation and staff that enabled this experiment.
We acknowledge support from the Deutsche Forschungsgemeinschaft (DFG) through SFB 1143 (Project No. 247310070) and the Würzburg-Dresden Cluster of Excellence on Complexity, Topology and Dynamics in Quantum Matter—ctd.qmat (EXC 2147, Project No. 390858490), as well as the support of the HLD at HZDR, member of the European Magnetic Field Laboratory (EMFL).
We acknowledge the European Synchrotron Radiation Facility (ESRF) for provision of synchrotron radiation facilities under proposal HC5433 on beamline ID32.
We acknowledge DESY (Hamburg, Germany), a member of the Helmholtz Association HGF, for the provision of experimental facilities.
Parts of this research were carried out at PETRA III.
Data was collected using beamlines P23 and P21.1 operated/provided by DESY Photon Science (or Helmholtz-Zentrum Hereon if applicable).
Beamtime was allocated for proposals R-20240670 EC and I-20250541 EC.

\medskip
\textit{Funding}
M.S. acknowledges support of Ministry of Science and Higher Education, Poland, under contracts 1/SOL/2021/2 and 2022/WK/13.
The work was supported by the Danish National Committee for Research Infrastructure (NUFI) through the ESS-Lighthouse Q-MAT, by the Danish Agency for Higher Education and Science through the instrument centre Danscatt, and by a research grant (Grant Agreement No. 35921) from VILLUM FONDEN.

\medskip
\textit{Data availability:} The data recorded at European XFEL~\cite{Yamamoto2024_XFEL006776} and ESRF~\cite{Martinelli2024_ESRF_HC5433} are available after the standard embargo period.